\documentclass[conference]{IEEEtran}
\IEEEoverridecommandlockouts
\usepackage{cite}
\usepackage{amsmath,amssymb,amsfonts}
\usepackage{algorithmic}
\usepackage{graphicx}
\usepackage{textcomp}
\usepackage{fancyhdr} 
\usepackage[table,xcdraw]{xcolor} 
\usepackage{colortbl}
\usepackage{subcaption}
\usepackage{multirow}
\usepackage{multicol}

\def\BibTeX{{\rm B\kern-.05em{\sc i\kern-.025em b}\kern-.08em
    T\kern-.1667em\lower.7ex\hbox{E}\kern-.125emX}}

\begin{document}

\title{SpecWav-Attack: Leveraging Spectrogram Resizing and Wav2Vec 2.0 for Attacking Anonymized Speech}


\author{\IEEEauthorblockN{1\textsuperscript{st} Yuqi Li}
\IEEEauthorblockA{\textit{Dept. of Large Language Model} \\
\textit{Qifu Technology}\\
Shanghai, China \\
liyuqi1-jk@360shuke.com}
\and
\IEEEauthorblockN{2\textsuperscript{nd} Yuanzhong Zheng}
\IEEEauthorblockA{\textit{Dept. of Large Language Model} \\
\textit{Qifu Technology}\\
Shanghai, China \\
zhengyuanzhong-jk@360shuke.com}
\and
\IEEEauthorblockN{3\textsuperscript{rd} Zhongtian Guo}
\IEEEauthorblockA{\textit{Dept. of Computer Science} \\
\textit{Fudan University}\\
Shanghai, China \\
guozt24@m.fudan.edu.cn}
\and
\IEEEauthorblockN{4\textsuperscript{th} Yaoxuan Wang}
\IEEEauthorblockA{\textit{Dept. of Large Language Model} \\
\textit{Qifu Technology}\\
Shanghai, China \\
wangyaoxuan-jk@360shuke.com}
\and
\IEEEauthorblockN{5\textsuperscript{th} Jianjun Yin*}
\IEEEauthorblockA{\textit{School Of Information Science And Technology} \\
\textit{\hspace{20mm} Fudan University}\\
\hspace{20mm} Shanghai, China \\
\hspace{20mm}yinjianjun@fudan.edu.cn}
\and
\IEEEauthorblockN{6\textsuperscript{th} Haojun Fei*}
\IEEEauthorblockA{\textit{Dept. of Large Language Model} \\
\textit{Qifu Technology}\\
Shanghai, China \\
feihaojun-jk@360shuke.com}
}

\maketitle

\begin{abstract}

This paper presents SpecWav-Attack, an adversarial model for detecting speakers in anonymized speech. It leverages Wav2Vec2 for feature extraction\cite{baevski2020wav2vec} and incorporates spectrogram resizing and incremental training for improved performance. Evaluated on librispeech-dev and librispeech-test, SpecWav-Attack outperforms conventional attacks, revealing vulnerabilities in anonymized speech systems and emphasizing the need for stronger defenses, benchmarked against the ICASSP 2025 Attacker Challenge\cite{tomashenko2024first}.

\end{abstract}


\section{Introduction}
This paper introduces SpecWav-Attack, a tailored adversarial model for attacking anonymized speech with a focus on Effective Equal Error Rate (EER). Using the ECAPA-TDNN architecture \cite{desplanques2020ecapa}, we integrate the Wav2Vec2 self-supervised model \cite{baevski2020wav2vec} to enrich speech representations, enhancing sensitivity to variations in anonymized data.

We apply a spectrogram resizing technique to the train-clean-360 dataset \cite{panayotov2015librispeech}, which boosts robustness and generalization. Evaluations on the librispeech-dev \cite{panayotov2015librispeech} and librispeech-test \cite{panayotov2015librispeech} datasets show that SpecWav-Attack outperforms traditional methods in both attack success rate and robustness.

Our results reveal vulnerabilities in speech anonymization and emphasize the need for more effective defenses in voice privacy, suggesting that advancements in robust countermeasures are crucial to mitigate adversarial risks.

\section{Methodology}
\subsection{Data Augmentation}
\begin{figure*}[htbp]
    \centering
    \includegraphics[width=\textwidth, height=0.4\textheight, trim=0 65 120 0, clip]{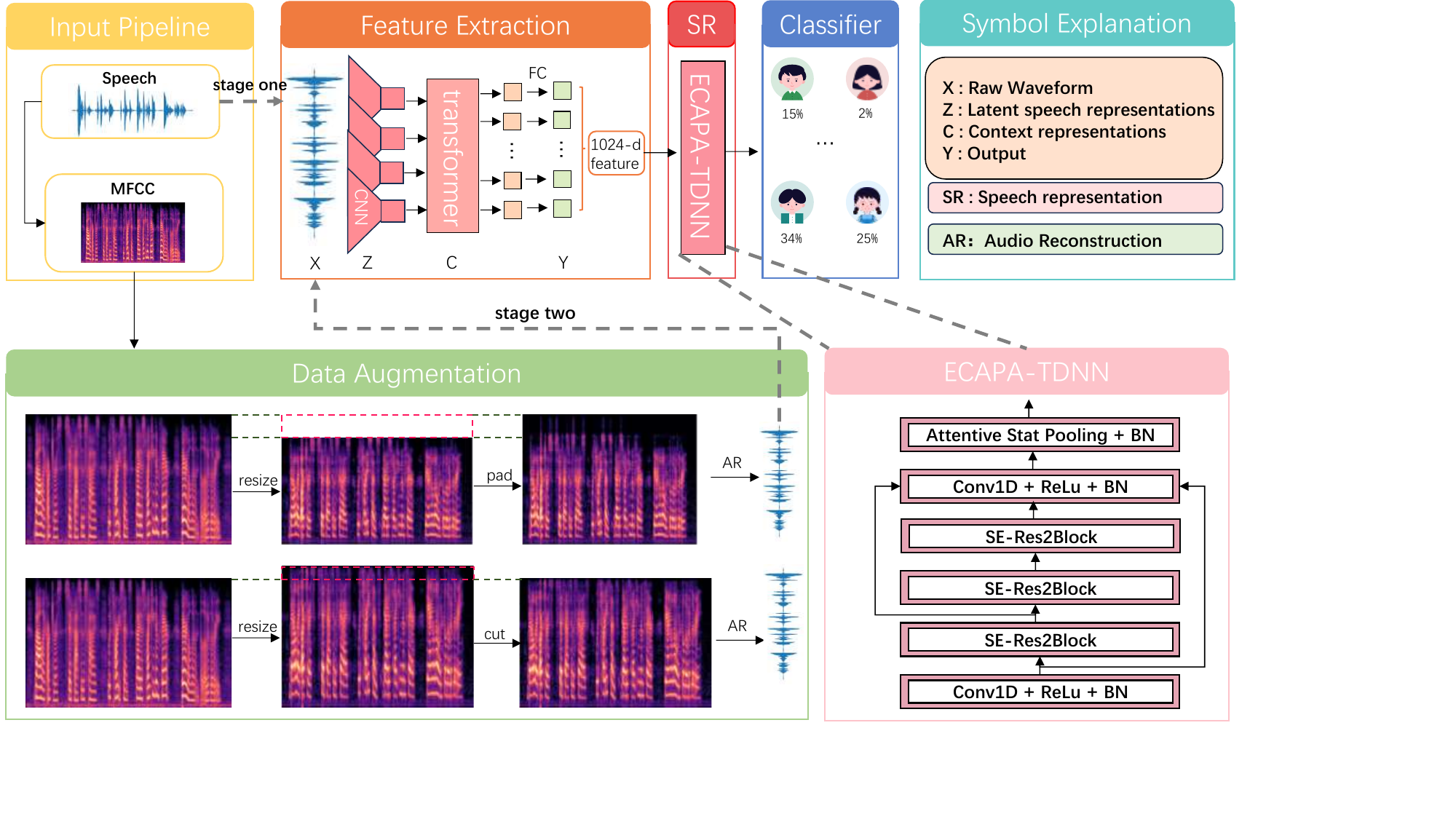}
    \caption{SpecWav-Attack Architecture: Stage one represents the first 10 epochs of training, and stage two represents 10-12 epochs of training. In this stage, the training set is replaced with the augmented dataset.}
    \label{fig:SpecWav-Attack Architecture}
\end{figure*}

We implement the SR-based data augmentation method proposed in Freevc\cite{li2023freevc}, which adjusts the Mel spectrogram vertically during preprocessing. As outlined in Fig. \ref{fig:SpecWav-Attack Architecture}, the augmentation process begins with the extraction of the Mel spectrogram \(x_{\text{mel}}\) from the source waveform \(y\). This is followed by vertical spectral resampling (SR) to produce a modified Mel spectrogram \(x_{\text{mel}}'\), which distorts speaker-specific features while retaining content-related information, thus enhancing model robustness. The process is completed by reconstructing the waveform \(y'\) from \(x_{\text{mel}}'\), thereby enhancing the model's adaptability to diverse speech inputs.

\subsection{Incremental Training}
\begin{figure}[htbp]
    \centering
    \begin{minipage}{0.48\linewidth}
        \centering
        \includegraphics[width=\linewidth]{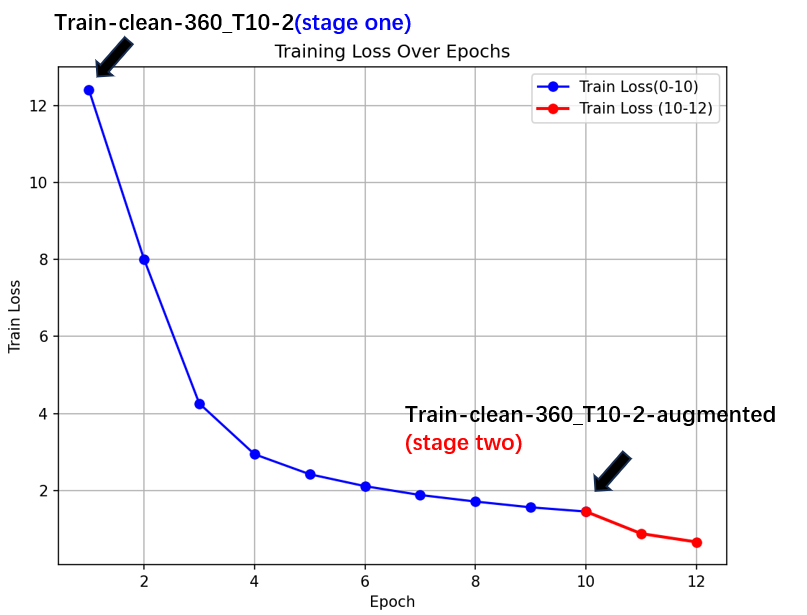}  
        \caption{Training loss changes with epoch}
        \label{fig:trian}
    \end{minipage}\hfill
    \begin{minipage}{0.48\linewidth}
        \centering
        \includegraphics[width=\linewidth]{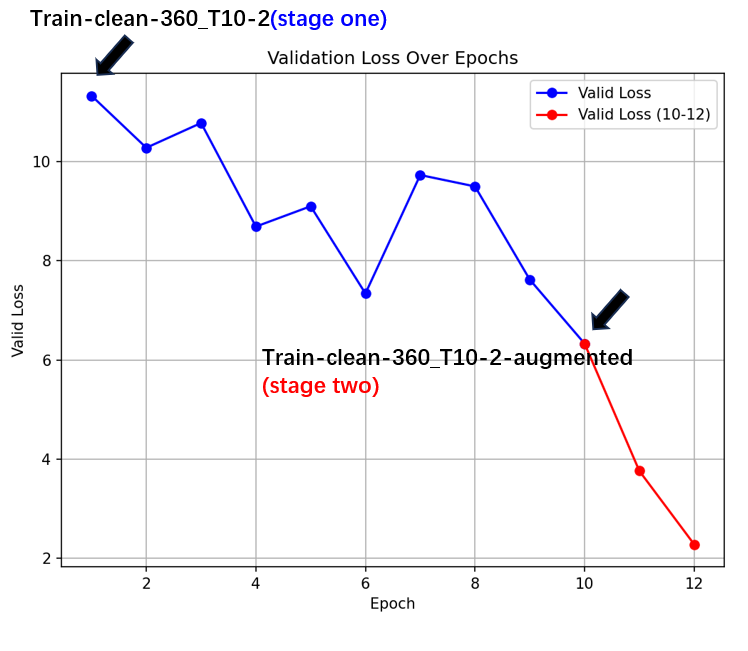}  
        \caption{Validating loss changes with epoch}
        \label{fig:valid}
    \end{minipage}
\end{figure}
\captionsetup[figure]{skip=2pt}  

\begin{table*}[htbp]
\centering
\begin{tabular}{|c|c|c|c|c|c|c|c|c|c|c|}
\hline
\multirow{2}{*}{Dataset} & \multirow{2}{*}{Gender} & \multicolumn{9}{c|}{EER(\%)} \\ \cline{3-11} 
                         &                        & Orig.  & T8-5     & T8-5(SW) & T10-2  & T10-2(SW) & T12-5  & T12-5(SW) & T25-1  & T25-1(SW) \\ \hline \hline
\multirow{2}{*}{LibriSpeech-dev} & female & 10.51 & 39.63 & 28.69 & 43.63 & 33.92 & 43.32 & 35.09 & 42.65 & 35.80 \\ 
                         & male   & 0.93  & 40.84 & 32.14 & 40.04 & 28.76 & 44.10 & 34.80 & 40.06 & 37.12 \\ \hline
\rowcolor{gray!20}  
\multicolumn{2}{|c|}{Average dev}  & 5.72 & 40.24 & 30.42  & 41.83 & 31.34 & 43.71 & 34.95 & 41.36 & 36.46 \\ \hline
\multirow{2}{*}{LibriSpeech-test}  & female & 8.76  & 42.50 & 28.09 & 41.97 & 25.91 & 43.61 & 34.85 & 42.34 & 35.58 \\ 
                         & male   & 0.42  & 40.05 & 29.63 & 29.85 & 27.17 & 41.88 & 34.52 & 41.92 & 35.19 \\ \hline
\rowcolor{gray!20} 
\multicolumn{2}{|c|}{Average eval}  & 4.59 & 41.28 & 28.86 & 40.36 & 26.54 & 42.75 & 34.69 & 41.35 & 36.39 \\ \hline
\end{tabular}
\caption{EER(Equal Error Rate) results for LibriSpeech-dev and LibriSpeech-test datasets,the SW(SpecWav) is the solution we proposed}
\label{tab:EER_results}
\end{table*}

Initially, a preliminary training phase is conducted for a certain number of epochs (10 epochs on the train-clean-360 dataset). After this, the model is further trained for 2 to 4 epochs on the train-clean-360-augmented dataset, building upon the previously trained model. This approach allows the model to gradually adapt to the data and the task, helping to avoid overfitting at the early stages of training.

\subsection{Feature Extraction}
In this work, we replace traditional fbank-based feature extraction with the self-supervised Wav2Vec 2.0 model, which generates 1024-dimensional embeddings from large unlabeled datasets to capture richer and more complex speech patterns. This model encodes both phonetic and speaker-independent features, enhancing performance in scenarios with limited labeled data. The improved feature extraction enhances the model's robustness and adaptability, making it ideal for tasks like anonymized speech recognition and other privacy-sensitive applications.

\section{Results}
Please refer to TABLE \ref{tab:EER_results}.

\section{Conclusion}

In summary, the SpecWav model significantly improves the Equal Error Rate (EER) over all baseline methods (T8-5, T10-2, T12-5, T25-1) on both the LibriSpeech-dev and LibriSpeech-test datasets. The improvements are particularly noticeable for T10-2, where a 13.82\% reduction in EER is observed. These results demonstrate the effectiveness of SpecWav in enhancing the performance of voice-based tasks by reducing EER across different configurations.


\vspace{12pt}

\end{document}